\documentstyle[12pt,epsf]{article}
\date{}
\begin{document}
\rightline{\bf CU-TP-954}
\vskip100pt
\bigskip

\begin{center}
{\Large\bf Small-x Physics, High Parton Densities\\

 and Parton Saturation in QCD\footnote{Lectures given at International Summer School ``Particle Production Spanning MeV and TeV
Energies,'' Nijmegen (Aug.8-20, 1999) and XVII Autumn School ``QCD: Perturbative or Nonperturbative?,'' Lisbon (Sept.29-Oct. 4, 
1999)}}

\vskip20pt
{A.H. Mueller\footnote{This work is sponsored in part by the Department of Energy, Grant DE-FG02-94-ER-40819}\\
Physics Department, Columbia University\\
New York, N.Y. 10027}
\end{center}

\begin{abstract}
Partons are defined as the quanta in a Fock space description of a hadron.  Gluon saturation is described in the Weizs\"{a}cker-Williams
approximation for a large nucleus.  The elements of DGLAP and BFKL evolution are given with the BFKL equation derived in a large-$N_c$ dipole
formalism.  A more general discussion of saturation is given in terms of a dipole scattering on a nucleon or nucleus.  Possible evidence for
saturation at HERA is discussed.
\end{abstract}
\section{Introduction:  Partons,classical fields and a simple picture of saturation}

In this section partons are defined as the quanta in the wavefunction of a hadron, or dressed quark, when quantization is done on the
light-cone and when light-cone gauge is used.  The relationship between this ``quantum'' description and a picture where gluons are viewed in
terms of a classical, or semiclassical, field in a hadron is then given.  Although the more classical description has limitations it also
provides a useful and intuitive way to picture many aspects of a high density parton system in terms of strong field configurations.  Gluon
saturation, in parton language, becomes a limit on the strength which the light-cone potential can reach, in the language of field
configurations.

\subsection{The field and gluon distribution in a quark}

Consider the wavefunction of a quark at order \ g\ in QCD.  The formula

\begin{equation}
\vert\psi_p) = N[\vert p> + \sum_{\lambda = \pm}\  \sum_{ c=1}^8 \int d^3k \psi_\lambda^c(k)\vert p-k, k(\lambda, c) >]
\end{equation}

\noindent is illustrated in Fig.1 and represents a dressed quark state in terms of a bare quark, the first term on the right-hand side of (1),
and a bare quark plus a gluon.  N  is a normalization factor, $\lambda = \pm$ are the polarizations of the virtual gluon, and $c = 1 \cdot \cdot
\cdot 8$ are the color indices of the gluon.  We have suppressed the color indices for the quark and instead view $\psi_\lambda^c$ as a
$3$\ x\ $3$ matrix in quark color space with the quark state being a 3-spinor in color space.  The integration in (1) is $d^3k=dk_+d^2\underline{k}$
with $k_\pm={1\over{\sqrt{2}}}(k_0\pm k_3).$

\begin{center}
\begin{figure}[htb]
\epsfbox{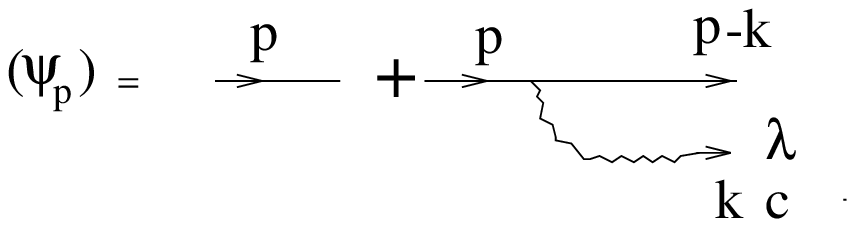}
\end{figure}
\center{FIG.1}
\end{center}

Recalling that in light-cone quantization momenta $P_+$ and $\underline{P}$ are kinematic while  $P_-$plays the role of the Hamiltonian, with
$P_-=\underline{P}^2/2P_+$ for an on-shell zero mass particle, $\psi$ can be written from light-cone perturbation theory as

\begin{equation}
\delta^3(p-p^\prime) \psi_\lambda^c(k) = {(p^\prime-k, k(\lambda,c)\vert g \int d^3x\bar{q}(x)\gamma_\mu{\lambda^c\over 2}q(x)A_\mu^c(x)\vert p)\over
(p^\prime -k)_-+k_--p_-}
\end{equation}

\noindent where $d^3x=d^2\underline{x}dx_-$ and where the free gluon field can be written as

\begin{equation}
A_\mu^c(x) = \sum_\lambda\int {d^3k\over {\sqrt{(2\pi)^32k_+}}}[\epsilon_\mu^\lambda(k) a_\lambda^c(k)
e^{i\underline{k}\cdot\underline{x}-ik_+x_--ik_-x_+}+h.c.]
\end{equation} 

\noindent with

\begin{equation}
[a_{\lambda^\prime}^{c^\prime}(k^\prime), a_\lambda^{c\dag}(k)] = \delta_{\lambda\lambda^\prime} \delta_{cc^\prime} \delta^3(k-k^\prime)
\end{equation}

\noindent and where $\delta^3(k-k^\prime)= \delta(\underline{k}-\underline{k}^\prime) \delta(k_+-k_+^\prime).$

It is useful to imagine the calculation done in a  frame where $p_+$ is large and $\underline{p}= 0$ in which case

\begin{equation}
k_-={\underline{k}^2\over 2k_+}, \ \ (p-k)_-={\underline{k}^2\over 2(p-k)_+}.
\end{equation}

\noindent In the soft gluon approximation (the small-x approximation) $k_+/p_+ << 1,$ and this implies that $k_->> (p-k)_-.$  Thus the
``energy'' of the softest particle, the gluon, dominates the energy denominator in (2).  Also, in $A_+=0$ light-cone gauge

\begin{equation}
\epsilon_\mu^\lambda(k) = (\epsilon_+^\lambda, \epsilon_-^\lambda, \underline{\epsilon}^\lambda) = (0,
{\underline{\epsilon}^\lambda\cdot\underline{k}\over k_+}, \underline{\epsilon}^\lambda)
\end{equation}

\noindent and, again in the soft gluon approximation, one need keep only the $\epsilon_-^\lambda$ term from (6) in (2) and (3).  Thus, one finds

\begin{equation}
\psi_\lambda^c(k) = ({\lambda^c\over 2}) 2g {(\underline{\epsilon}^\lambda)^\ast\cdot\underline{k}\over \underline{k}^2} \ {1\over
{\sqrt{2\pi)^32k_+}}}.
\end{equation}
\vskip15pt
\noindent {\bf Problem 1(E):} Use (2) along with (3)-(6) and $\tilde{U}(p-k) \gamma_\mu U(p) \approx 2p_+g_{\mu_-}$ to derive (7).  
 \vskip10pt
 Now define the gluon distribution of the quark by

\begin{equation}
 x G_q(x,Q^2) = \int
\sum_{\lambda,c}d^2\underline{k}\Theta(Q^2-\underline{k}^2)x\delta (x-k_+/p_+)(\psi_p\vert\ a_\lambda^{c^\dag}(k)a_\lambda^c(k)\vert\psi_p)
\end{equation}

\noindent which immediately gives

\begin{displaymath}
x G_q(x,Q^2) =  \sum_{c,\lambda}\int d^3k\  x\  \delta(x-k_+/p_+)\Theta(Q^2-\underline{k}^2)\psi_\lambda^{\dag c}(k)\psi_\lambda^c(k)
\end{displaymath}
\begin{equation}
 = \sum_c {\lambda^c\over 2} {\lambda^c\over 2} 
\int {4g^2\over (2\pi)^3} \ {d^2\underline{k}\over \underline{k}^2}\ {dk_+\over 2k_+}\ x
\delta(x-k_+/p_+) \Theta(Q^2-\underline{k}^2).
\end{equation}

\noindent Introducing an infrared cutoff, $\mu^2,$ for the transverse momentum integral and using $\sum_c{\lambda^c\over 2}\ {\lambda^c\over 2}\
=\ C_F\ = {N_c^2-1\over 2N_c}$ and $g^2=4\pi\alpha$ one finally obtains

\begin{equation}
xG_q(x,Q^2) = {\alpha C_F\over \pi}\ \ell n\ Q^2/\mu^2 .
\end{equation}

\noindent Equation (8) is a natural and intuitive definition of the gluon distribution as it clearly corresponds to the number of gluons per
unit of \ x\ in the wavefunction having transverse momentum less than  Q.  The free integration of $d^2\underline{k}$ in (8) for
$\underline{k}^2 < Q^2$ can be viewed as determining that the gluon, and the quark from which it came, are point-like and localized down to a
transverse size $\Delta x_\perp=2/Q.$  We note that if one takes $xG(x,Q^2) = 3xG_q(x,Q^2) = {3\alpha C_F\over \pi} \ell n Q^2/\mu^2,$ one obtains a result
for the proton which is not unreasonable phenomenologically for $x \sim 10^{-2}-10^{-1}$ and moderate $Q^2$ if  $\mu$ is taken to be 100 MeV.

One can associate a ``classical'' field with the gluon in the quark according to

\begin{equation}
A_\mu^{c\ell}(x) = \int\ d^3p^\prime (\psi_{p^\prime}\vert A_\mu^c(x)\vert \psi_p)
\end{equation}

\noindent or, for $i=1,2,$

\begin{equation}
A_i^{c(c\ell)}(x) = \int {d^3k\over (2\pi)^3}\ e^{- ik\cdot x}({\lambda^c\over 2}) {gk_i\over \underline{k}^2k_+}.
\end{equation}

\noindent It is then natural to define the ``classical'' field in momentum space to be
 
\begin{equation}
A_i^{c(c\ell)}(k) = {\lambda^c\over 2} \ {gk_i\over \underline{k}^2 k_+}.
\end{equation}
\vskip15pt
\noindent{\bf Problem 2(M):}  Take ${1\over k_+}\ = {1\over k_+-i\epsilon}$ in (12) and show that 
\vskip10pt
\begin{displaymath}
A_i^{c(c\ell)}(x) = - q({\lambda^c\over 2}) {x_i\Theta(-x_-)\over 2\pi \underline{x}^2}\  {\rm and} \ F_{+i}^{c(c\ell)}=
{\partial\over \partial x_-} A_i^{c(c\ell)} = g({\lambda^c\over 2}) {x_i\over 2\pi\underline{x}^2} \delta(x_-).
\end{displaymath}
\vskip20pt
\noindent {\bf Problem 3(M-H):} Derive (12) using the Feynman graph in Fig.2 along with the light-cone gauge propagator
$D_{\mu\nu}(k) =\ {-i\over k^2+ i\epsilon} [g_{\mu\nu} - {\eta_\mu k_\nu\over k_++i\epsilon}\ -\ {\eta_\nu k_\mu\over k_+-i\epsilon}]\ {\rm
where}\ \ \  \eta\cdot V = V_+$ for any vector $V_\mu.$

\vskip10pt
\noindent {\bf Problem 4(E):}  Show that $G_q,$ as given in (9), can be obtained from 
\vskip10pt
\begin{displaymath}
G_q(x,Q^2) = \int {d^3k\over (2\pi)^3}
\Theta(Q^2-\underline{k}^2) 2k_+\delta(x-k_+/p_+) \sum_{c,i}\vert A_i^{c(c\ell)}(k)\vert^2.
\end{displaymath}

\begin{center}
\begin{figure}[htb]
\epsfbox[0 0 193 180]{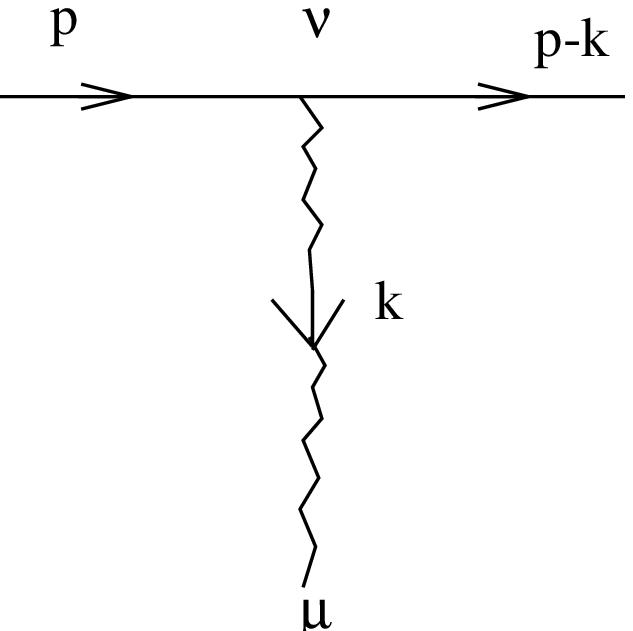}
\end{figure}
\center{FIG.2}
\end{center}
   The $i\epsilon$ prescriptions in problems 2 and 3 may seem a little unusual.  We shall have more to say about this later on.  The fact
that the $i\epsilon$'s are different in the two $k_+$ denominators in the propagator in problem 3, however, is necessary for consistency.  If we
orient the gluon momentum according to the arrow in Fig.2 then the $-i\epsilon$ corresponds to the case that the $\eta$ carries the index at the
tail of the arrow.  This uniquely prescribes how the  $i\epsilon$'s are to be given.  Finally, problem 4 shows that the gluon distribution can
be written in terms of the ``classical'' gauge potential a quantity intimately connected with the $dx_--$integral of the field strength as is
clear from problem 2.  The classical field we have defined here for gluons is identical, except for the color matrix ${\lambda^2\over 2},$
to that which one naturally defines for photons in quantum electrodynamics.

\subsection{The gluon field and distribution in a proton}

A quark can emit, or absorb, a gluon and remain a quark, however, a proton cannot.  Thus we generalize the idea of a classical gluon field in a
proton\cite{Yu}, and instead of the result given in problem 2  we take

\begin{equation}
A_i^{c(c\ell)}(x) =\ -\ g\int d^2bdb_-{(x-b)_i\over 2\pi(\underline{x}-\underline{b})^2} \Theta(b_--x_-) \hat{\rho}^c(\underline{b},b_-)
\end{equation}

\noindent where $\hat{\rho}$ is the quark color charge density operator for the proton.  It is convenient to introduce a matrix notation for
the gauge potential where $A_\mu\  =  \sum_c{\lambda^c\over 2}\ A_\mu^c$ and $A_\mu^c=2 tr[{\lambda^c\over 2}A].$  Thus in matrix notation

\begin{equation} 
A_i^{c\ell}(x) = - g\int d^3b{(x-b_i)\over 2\pi(\underline{b}-\underline{x})^2}\Theta(b_--x_-)\hat{\rho}(\underline{b},b_-)
\end{equation}

\noindent and

\begin{equation}
A_i^{c\ell}(k)= {gk_i\over \underline{k}^2(k_+-i\epsilon)}\hat{\rho}(k)
\end{equation}

\noindent where

\begin{equation}
\hat{\rho}(k) = \int d^2bdb_- e^{-i\underline{k}\cdot\underline{b}+ik_+b_-}\hat{\rho}(b).
\end{equation}

\noindent The integration over\   $b$\  in (14) - (17) goes over the volume of the proton with $\underline{b}$ the usual transverse coordinate
and $b_-$ the light-cone analog of the usual z-coordinate.  We shall only use (15), or (16), for transverse wavelengths much smaller than the
radius of the proton in which case we may make $\hat{\rho}$ more precise by writing\cite{Yu}

\begin{equation}
<\hat{\rho}^c(b)\hat{\rho}^{c\prime}(b^\prime)> = {\delta_{cc^\prime}\over 2N_c}
\delta(\underline{b}-\underline{b}^\prime)\delta(b_-  -b_-^\prime) \rho(\underline{b},b_-)
\end{equation}

\noindent for the charge-charge correlator in the proton where $\rho(\underline{b}, b_-)$ is the quark number density normalized according to

\begin{equation}
\int d^2bdb_-\rho(b) = N_c.
\end{equation}
\noindent Of course (18) simply reflects the fact that for short transverse spatial separations only individual quarks are effective while (19)
reflects the fact that we are working in a valence quark model\cite{Mc}, and that we do not consider the evolution in the proton which would
produce either a quark sea or gluons additional to those which will couple to $\hat{\rho}$ through (15) and (16).  That is we shall only
allow two gluons to be emitted or absorbed in a proton.  Finally, when $x_-$ is sufficiently negative

\begin{equation}
{2\over \pi} tr \int d^2b < A_i^{c\ell}(\underline{x} + \underline{b}, x_-) A_i^{c\ell}(\underline{b}, x_-) = xG(x,Q^2)
\end{equation}

\noindent with $Q^2$ identified with $4/\underline{x}^2.$  Equation (20) shows that we can recover the gluon distribution in a nucleon, as given
by the independent gluon distributions of the valence quarks, in terms of the ``classical'' gluon field we defined in (14).
\vskip 15pt
\noindent{\bf Problem 5(E): Derive (20)}
\vskip15pt
\subsection{Non-Abelian Weizs\"acker-Williams field and gluon saturation}

Now we wish to find $A_\mu^{(c\ell)}$ for a large nucleus.  We suppose that $A_\mu^{(c\ell)}$ is small for
a nucleon, but for a large nucleus $A_\mu^{(c\ell)}$, the light -cone gauge potential, will be large and
non-linearities will have to be taken into account.  It is these non-linearities which lead to the idea of
gluon saturation.  Our approach is the following.  In a nucleus it is well-known that interactions between the
nucleons are short range and, to a good approximation, these interactions are weak enough, because of the
diluteness of nuclear matter, that they may be neglected altogether in scattering hadrons and real and
virtual photons on nuclei.  If we express the gluon field of a nucleon in a covariant gauge then the
gluon field will decrease reasonably rapidly outside a nucleon, even without having a model of confinement, due
to the color neutrality of the nucleon.  Thus we are going to assume that the ``classical'' gluon field of a
nucleus is additive in the nucleons in covariant gauge and then transform that result to light-cone gauge
where it no longer will be additive, but where the partonic interpretation will be manifest\cite{Yu}.  We
begin by noting that in a covariant, Lorentz, gauge the classical field of a quark becomes
                                  
\begin{equation}
A_+^{^\prime c(c\ell)}(x) = - g({\lambda^c\over 2}){\delta(x_-)\over 2\pi}\ \ell n[\vert\underline{x}\vert\mu];\ A_i^\prime=A_-^\prime=0
\end{equation}

\noindent instead of that given in problem 2.
\bigskip
\noindent {\bf Problem 6(E):}  Show that the same $F_{+i}$ as given in Problem 2 can be obtained from $F_{+_i}^{c(c\ell)}(x) = -
\bigtriangledown_iA_+^{^\prime c(c\ell)}(x).$
\bigskip

The Lorentz gauge potential for a nucleon is then given by

\begin{equation}
A_+^{^\prime(c\ell)}(\underline{x}, x_-) = - g\int{d^2b\over 2\pi} \ell n[\vert\underline{x}-\underline{b}\vert\mu]\hat{\rho}(\underline{b},x_-)
\end{equation}

\noindent with the $\hat{\rho}(b) = \sum_c\ {\lambda^c\over 2}\ \hat{\rho}^c(b)$ the same as in (15) and (18).  The infrared cutoff, $\mu,$ in
(21) and (22) will disappear when physical quantities are evaluated.  Now we take the Lorentz gauge potential for a large nucleus simply by
adding the contributions from the individual nucleons as given in (22).  In order to transform this result to light-cone gauge we use the
general formula for changing gauges in QCD

\begin{equation}
A_\mu(x) = SA_\mu^\prime(x) S^{-1}-i/g(\partial_\mu S)S^{-1}
\end{equation}

\noindent where $S$ is a (space-time dependent) unitary transformation to be determined.  In order that $A_\mu$ correspond to a light-cone
gauge potential we require $A_+=0$ in (23) which gives

\begin{equation}
{\partial\over \partial x_-} S = - ig SA_+^\prime.
\end{equation}

\noindent Equation (24) is easily solved as

\begin{equation}
S(\underline{x}, x_-) = P\  exp\{-ig \int_{-\infty}^{x_-} dx_-^\prime A_+^\prime(\underline{x}, x_-^\prime)\}
\end{equation}

\noindent where the\  $P$\ symbol in (25) indicates that one should order the matrices $A_+^\prime$ so that $A_+^\prime(\underline{x},
x_-^\prime)$ comes to the right (left) of $A_+^\prime(\underline{x}, x_-^{\prime\prime})$ in case $x_-^\prime$ is greater (less) than
$x_-^{\prime\prime}.$  The choice of writing the path integral from $x_-^\prime =\ -\ \infty$ to $x_-^\prime = x_-$ corresponds to the choice
of $i\epsilon's$ in the light-cone denominators in problems 2 and  3 and in (16).

Now that $S$ has been determined in terms of the Lorentz gauge potential we go back to (23) to get $A_i^{(c\ell)}.$  The first term on the
right-hand side of (23) is not large since it is just a unitary transformation of the Lorentz gauge potential.  The large term is

\begin{displaymath}
A_i^{(c\ell)}(\underline{x}, x_-) =\ -\ {i\over g}\ (\partial_iS) S^{-1}
\end{displaymath}

\noindent leading to\cite{Yu}

\begin{equation}
A_i^{(c\ell)}(\underline{x}, x_-) =\ - \int_{-\infty}^{x_-} dx_-^\prime S(\underline{x}, x_-^\prime) \partial_iA_+^\prime(\underline{x},
x_-^\prime) S^{-1}(\underline{x}, x_-^\prime)
\end{equation}

\noindent where $S$ is given by (25) with $A_+^\prime$ given by (22).

The evaluation of (26) is a bit technical and we just outline the steps here\cite{V.K}.  We begin by assuming our ``nucleus'' is just a thin
slice in the $z(x_-)$ direction.  Then, later, we will assemble the various slices to get the total result for spherical nucleus.  Then for the
thin slice
\vfill
\begin{displaymath}
S(\underline{x}, x_-^\prime)\partial_iA_+^{\prime(c\ell)}(\underline{x}, x_-^\prime)S^{-1}(\underline{x}, x_-^\prime) 
\end{displaymath}
\begin{displaymath}
= - g\int{d^2b\over
2\pi}\partial_i\ell n[\vert\underline{x}-\underline{b}\vert \mu]\hat{\rho}^c(\underline{b},x_-^\prime)
\{{\lambda^c\over 2}\left[1-{g^4\over 4}\int{d^2\underline{b}^\prime d b_-^\prime b_-\over (2\pi)^2}
\rho(\underline{b}^\prime,b_-^\prime)\ell n^2[\vert\underline{x}-\underline{b}^\prime\vert\mu]\right]
\end{displaymath}
\begin{equation}
-g^2fcde{\lambda^e\over 2}
\int{d^2\underline{b}^\prime db_-^\prime\over 2\pi} \ell
n[\vert\underline{x}-\underline{b}^\prime\vert\mu]
\hat{\rho}^d(\underline{b}^\prime,b_-^\prime)\}.
\end{equation}

\noindent In arriving at (27) we have anticipated our next step and used (18) for the quadratic
term in $\hat{\rho}$ which leads to the $\rho \ell n^2$ term in (27) with $\rho(\underline{b}^\prime, b_-^\prime)$ the quark number density
in                                                                                                                            the nucleus
normalized according to
$\int\rho d^3b = N_cA.$  For a thin slice of nuclear matter we have kept only quadratic terms and linear terms in $\hat{\rho}$ in (27).  Now it
is straightforward to find, still for a thin slice of nuclear matter,

\begin{displaymath}
{2\over \pi} \int d^2b tr <A_i^{c\ell}(\underline{b}+\underline{x})A_i^{c\ell}(\underline{b})> = \int dx_-d^2b xG(x,Q^2)\rho(\underline{b},x_-)
\cdot
\end{displaymath}
\begin{equation}
\cdot\ \left\{1-\int_{-\infty}^{x_-}d b_-{\pi^2\underline{x}^2\alpha N_c\over N_c^2-1} \rho(\underline{b},b_-) xG(x, Q^2)\right\}
\end{equation}

\noindent  with $xG(x, Q^2) = N_c{\alpha C_F\over \pi} \ell n Q^2/\mu^2,$ the gluon distribution in a nucleon and where now
$\rho(\underline{b}, b_-)$ is the normal nucleon number density in the nucleus a factor of $1/N_c$ times the $\rho$ which appears in (27).  It
is now straightforward to go to a thick nucleus by simply exponentiating the second term in brackets on the right-hand side of (19).  Thus if we
call

\begin{equation}
\tilde{N}(\underline{x}^2) = {2\over \pi} tr \int d^2b <A_i^{c\ell}(\underline{b}+ \underline{x}) A_i^{c\ell}(\underline{b})> 
\end{equation}

\noindent then\cite{J.Ja,V.K}

\begin{equation}
\tilde{N}(\underline{x}^2) = \int d^2b {N_c^2-1\over \pi^2\alpha N_c\ \underline{x}^2} [1-e^{-\underline{x}^2Q_s^2/4}]
\end{equation}

\noindent where the saturation momentum, $Q_s$, is given by

\begin{equation}
Q_s^2 = {8\pi^2\alpha N_c\over N_c^2-1}{\sqrt{R^2-\underline{b}^2}}\ \rho xG(x,Q^2).
\end{equation}

\noindent{\bf Problem 7(M-H):}  Derive (27) and (28), from (18), (22), (25) and (26).  The derivation of (28) from (18) and (27) is not hard
but requires good bookkeeping skills along with the result of Problem 8.
\vskip 15pt
\noindent{\bf Problem 8(H):} Show that for small $\underline{x}^2$

\begin{displaymath}
\int d^2\underline{b}^\prime[\ell n\vert(\underline{b}+\underline{x}-\underline{b}^\prime)\mu\vert\cdot\ell n
\vert(\underline{b}-\underline{b}^\prime)\mu\vert-{1\over 2} \ell n^2\vert(\underline{b}-\underline{b}^\prime)\mu\vert-{1\over 2}\ell
n^2\vert(\underline{b}-\underline{b}^\prime +\underline{x})\mu\vert]\rho(\underline{b}^\prime,b_-)
\end{displaymath}
\begin{displaymath}
=\ -\ {\pi\over 4} \underline{x}^2 \rho(\underline{b}, b_-) \ell n 1/\underline{x}^2.
\end{displaymath}

\subsection{Interpreting saturation in the McLerran-Venugopalan model\cite{Mc}}

It is tempting to identify $\tilde{N}(\underline{x}^2)$ with $ xG_A(x,Q^2)$ as was done in (20) for the nucleon.  This identification is
correct if $Q^2= 4/\underline{x}^2$ lies far above $Q_s^2,$ but it is not correct for $Q^2$ below $Q_s^2.$  What we have actually been
calculating is the number density of gluons in the nuclear Fock space wavefunction.  So long as $Q^2/Q_s^2$ is much greater than one this number
density agrees with that measured in deep inelastic scattering.  However, the gluons in the nuclear wavefunction have been rearranged in
transverse momentum compared to what a simple superposition of nucleon gluon densities would suggest and at  the semiclassical level we are now
working the gluon distribution as defined by the operator product expansion, and as measured in deep inelastic scattering, is still
additive.

Thus

\begin{equation}
{dxG\over d^2\ell} = \int {d^2 x\over 4\pi^2} e^{i\underline{\ell}\cdot \underline{x}}\tilde{N}(\underline{x}^2)
\end{equation}

\noindent is the number of gluons per unit transverse momentum in the nuclear wavefunction while

\begin{equation}
{dxG\over d^2bd^2\ell} = \int {d^2x\over 4\pi^4}\ {N_c^2-1\over \alpha N_c\ \underline{x}^2} (1-e^{-\underline{x}^2Q_s^2/4})
e^{i\underline{\ell}\cdot\underline{x}}
\end{equation}

\noindent is the number of gluons per unit of transverse phase space in the wavefunction of the nucleus.
\vskip15pt
\noindent{\bf Problem 9(M):}  Show that the total number of gluons in the nucleon wavefunction, the integral of (33) over all  $b$ and $\ell$,
is simply the sum of the number of gluons in each of the nucleon wavefunctions even though the distribution of gluons in $\underline{\ell}$ has
been modified from a simple additive rule.  In particular,this means that there is no shadowing in the McLerran-Venugopalan model.

The result in Problem 9 shows that saturation and nuclear shadowing are very different phenomena.  Saturation is apparent in\cite
{J.Ja}

\begin{equation}
{1\over N_c^2-1} tr <A_i^{c\ell}(\underline{b} + \underline{x}) A_i^{c\ell}(\underline{b})> = {1\over 2\pi\alpha
N_c\underline{x}^2}(1-e^{-\underline{x}^2Q_s^2/4})
\end{equation} 

\noindent which follows from (29) and (30) and which says that the light-cone gauge potential never becomes larger than

\begin{equation}
{\sqrt{\underline{x}^2}} A_i^{c(c\ell)} \sim {1\over{\sqrt{4\pi\alpha N_c}}}
\end{equation}

\noindent the ``ultimate'' value of the potential allowed before gluonic ``interactions'' move gluons from lower transverse momenta to higher
transverse momenta.  We shall come back to discuss saturation in a more general context, and especially for quark densities, in the third
chapter of these lectures.

\section{DGLAP and BFKL evolution}

Dokshitzer, Gribov, Lipatov, Altarelli and Parisi (DGLAP)\cite{L.Do,Gri,G.Al} evolution and Balitsky, Fadin, Kuraev and Lipatov
(BFKL)\cite{E.A.K,Ya.Ya} evolution tell how to go from a (given) coarse-grained description of partons in the wavefunction to a more
fine-grained description.  In the case of DGLAP evolution, which is equivalent to the usual renormalization group evolution, one passes from a
coarse-grained description in $\Delta x_\perp$ (or in $Q$) to a more fine-grained description in $\Delta x_\perp$(or in $Q$), while in BFKL
evolution one moves from a coarse-grained description in $\Delta t$ (or in $x$) to a more fine-grained, shorter time, description.

\subsection{DGLAP evolution in the leading logarithnmic approximation}

Recall, from (9) and (10) we found an expression for $x G_q(x, Q^2).$  Clearly this expression obeys the equation

\begin{equation}
Q^2 {\partial\over \partial Q^2} x G_q(x, Q^2) = {\alpha C_F\over \pi}.
\end{equation}

\noindent If we interpret (35) as telling us how many gluons come from a single quark in a given small transverse momentum bin (see (9)) then we
can generalize this equation easily to the general distribution of gluons in a proton according to

\begin{equation}
Q^2 {\partial\over \partial Q^2} xG(x,Q^2) = {\alpha C_F\over \pi} \int_x^1 {dx^\prime\over x^\prime} x^\prime q(x^\prime,Q^2) + {\alpha
N_c\over
\pi} \int_x^1 {dx^\prime\over x^\prime} x^\prime G(x^\prime,Q^2).
\end{equation}

\noindent In the first term on the right-hand side of (37) the ${\alpha C_F\over \pi}$ factor is exactly as in (36) while the factor
multiplying ${\alpha C_F\over \pi}$ is the number of quarks (and antiquarks) having longitudinal momentum greater than the gluon emitted.  The
second term on the right-hand side of (37) corresponds to emission of the ``observed'' gluon off higher longitudinal momentum gluons and
$N_c=C_A$ naturally replaces $C_F.$
\vskip15pt
\noindent{\bf Problem 10(H):} Drop the $x^\prime q(x^\prime,Q^2)$ term in (37) and solve for $xG(x,Q^2)$ in the small-x and large-$Q^2$ limit,
in the leading double logarithmic approximation.  First suppose $\alpha$ is a constant.  Then, if you can, solve when 
 $\alpha = {1\over b \ell n Q^2/\mu^2}$ with $b = {11 N_c-2N_f\over 12\pi}.$

\subsection{The diple picture of BFKL evolution[10-12]}
\subsubsection{Lowest order}

At lowest order the cross section for heavy onium-heavy onium scattering is given in terms of the amplitude illustrated in Fig.3,
where the lines $P-k_1$ and $P^\prime-k_1^\prime$ are heavy quarks and the lines $k_1$ and $k_1^\prime$ are heavy antiquarks.

\begin{center}
\begin{figure}[htb]
\epsfbox{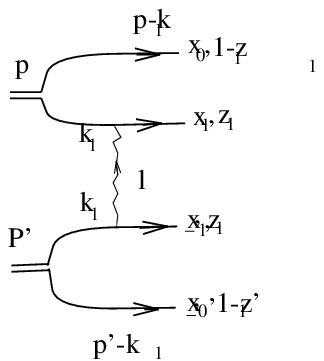}
\end{figure}
\center{FIG.3}
\end{center}

We find it convenient to describe the onium wavefunctions using transverse coordinate space variables, $\underline{x}_0^\prime, \underline{x}_1^\prime,
\underline{x}_0^\prime\underline{x}_1^\prime,$ and longitudinal momentum fractions,
$z_1, 1-x_1, z_1^\prime, 1-z_1^\prime,$ as indicated in Fig.3. The coordinate and momentum space wavefunctions are related by

\begin{equation}
\psi^{(0)}(x_{01}, z_1) = \int {d^2k_1\over (2\pi)^2} e^{i\underline{k}_1\cdot\underline{x}_{01}}\psi^{(0)}(k_1,z_1),
\end{equation}

\noindent where $\underline{x}_{01} = \underline{x}_1 - \underline{x}_0$ and the superscript $(0)$ indicates a lowest order (no soft gluons)
wavefunction.  In terms of $\Phi^{(0)}=\vert \psi^{(0)}\vert^2$ the lowest order cross section is

\begin{equation}
\sigma^{(0)}\ =\ \int d^2x_{01}\int_0^1 dz_1\Phi^{(0)}(x_{01},z_1) \int d^2x_{01}^\prime\int_0^1
dz_1^\prime\Phi^{(0)}(x_{01}^\prime,z_1^\prime)\sigma_{dd}(x_{01},x_{01}^\prime),
\end{equation}

\noindent where $\sigma_{dd}$ is the dipole-dipole scattering cross section

\begin{equation}
\sigma_{dd}(x,x^\prime) = 2\pi\alpha^2x_<^2 \left(1 + \ell n{x_>\over x_<}\right)
\end{equation}

\noindent with $x_>$ the greater of $x$ and $x^\prime$ and with $x_<$ the lesser.  Our normalization is such that

\begin{equation}
\int d^2x \int_0^1 dz\Phi^{(0)}(x,z) = 1.
\end{equation}\\

\noindent{\bf Problem{11(H):}}  Show

\begin{displaymath}
\sigma_{dd}(x_{01}, x_{01}^\prime) = \alpha^2 \int {d^2\ell\over [\underline{\ell}]^2}
(2-e^{i\underline{\ell}\cdot\underline{x}_{01}}-e^{-i\underline{\ell}\cdot\underline{x}_{01}})
(2-e^{i\underline{\ell}\cdot\underline{x}_{01}^\prime}-e^{-i\underline{\ell}\cdot\underline{x}_{01}^\prime})
\end{displaymath}

\noindent and use

\begin{displaymath}
\int_0^\infty{d\ell\over \ell^3}(1-J_0(\ell  x))(1-J_0(\ell x^\prime)) = {1\over 4} x_<^2\left(1 + 1n{x_>\over x_<}\right)
\end{displaymath}

\noindent to get (40) when orientations of $\underline{x}_{01}$ and $\underline{x}_{01}^\prime$ are averaged over.

\subsubsection{One soft gluon in the wavefunction}

The onium wavefunction having a single soft gluon, along with the heavy quark-antiquark, is calculated from the graphs in Fig.4, where softness
means $z_2<<1.$  In the large $N_c$ limit the emission of a gluon changes a dipole into two dipoles.  The original dipole, the heavy
quark-antiquark pair, becomes two dipoles consisting of the heavy quark and the antiquark part of the gluon making up the first dipole and the quark part of
the gluon along with the heavy antiquark making up the second dipole. We illustrate the sum of the two graphs in Fig.4 by the single graph of Fig.5 where
the dipole structure is emphasized.

\begin{center}
\begin{figure}[htb]
\epsfbox{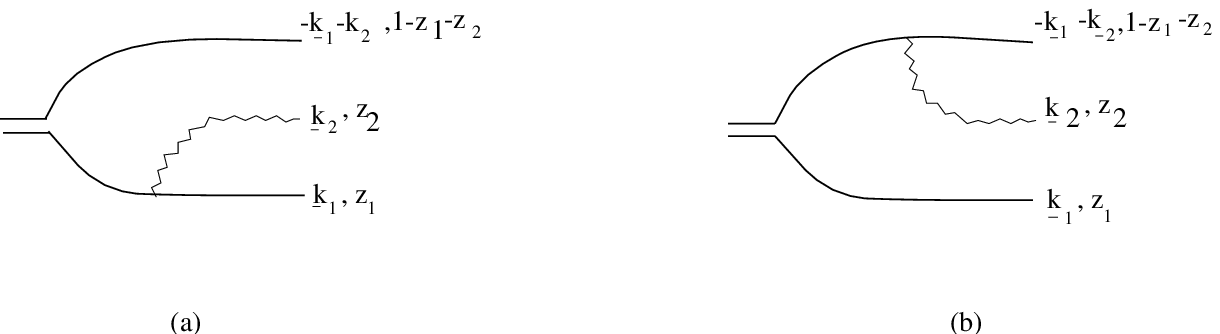}
\end{figure}
\center{FIG.4}
\end{center}
In order to derive the BFKL equation one needs only to calculate the graphs of Fig.4 in light-cone gauge where the polarization of the gluon
is

\begin{equation}
\epsilon_\mu^\lambda(k) = (\epsilon_+^\lambda, \epsilon_-^\lambda,\underline{\epsilon}^\lambda) = (0, {\underline{\epsilon}^\lambda\cdot
\underline{k}\over k_+}, \underline{\epsilon}^\lambda)
\end{equation}

\noindent In the soft gluon approximation, which corresponds to the usual leading logarithmic approximation in which BFKL dynamics is
formulated, one need only keep $\epsilon_-^\lambda$ in (42) because the $1/k_+$ is a large quantity.  Using the fact that the $\epsilon_-$
polarization couples to the classical current of a high momentum quark or antiqark we can immediately write down the contribution of the
graphs in Fig.4 to the wavefunction of the onium as

\begin{equation}
\psi^{(1)}(k_1, k_2, z_1,z_2) = gT^a{1\over \underline{k}_2^2/2k_{2+}} {\underline{\epsilon}^\lambda\cdot\underline{k}_2\over k_{2+}}
(\psi^{(0)}(k_1+k_2,z) - \psi^{(0)}(k_1,z_1)),
\end{equation}

\begin{center}
\begin{figure}
\epsfbox{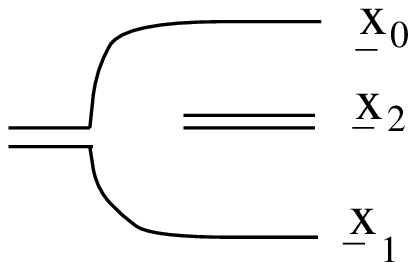}
\center{FIG.5}
\end{figure}
\end{center}

\noindent where $T^a$ is the color matrix for a gluon of color  $a,$  and $[\underline{k}_2^2/2k_{2+}]^{-1}$ is the energy denominator which is
dominated by the soft gluon.  Going to transverse coordinate space by

\begin{equation}
\psi^{(1)}(x_{02}, x_{21}, z_1, z_2) = \int {d^2k_1d^2k_2\over (2\pi)^4} e^{i\underline{k}_1\cdot\underline{x}_{01} +
i\underline{k}_2\underline{x}_{02}}\psi^{(1)}(k_1,k_2,z_1,z_2)
\end{equation}

\noindent one finds, using (43),

\begin{equation}
\psi^{(1)}(x_{02}, x_{21}, z_1, z_2) = \psi^{(0)}(x_{01}, z_1)\  {igT^a\over \pi}\  \left({\underline{x}_{02}\over x_{02}^2} +
{\underline{x}_{21}\over x_{21}}\right).
\end{equation}

\noindent Squaring and taking a  trace over colors gives

\begin{equation}
\Phi^{(1)} = \Phi^{(0)}(x_{01}, z_1)\ {2\alpha N_c\over \pi}\ {x_{01}^2\over x_{12}^2 x_{02}^2}.
\end{equation}

\noindent In obtaining (46) we have taken $C_F=N_c/2$ in the large $N_c$ limit.
\vskip15pt
\noindent{\bf Problem 12(E):}  Show

\begin{displaymath}
\sum_\lambda\left[\underline{\epsilon}^\lambda\cdot \left({\underline{x}_{02}\over x_{02}^2} + {\underline{x}_{21}\over
x_{21}^2}\right)\right]^2 = {x_{01}^2\over x_{12}^2 x_{02}^2}.
\end{displaymath}

\vskip15pt
\noindent Including a phase space factor

\begin{displaymath}
d\Omega = {d^2x_2\over 2\pi}\ {dz_2\over 2z_2}
\end{displaymath}

\noindent one arrives at the factor

\begin{equation}
{1\over\Phi^{(0)}} \Phi^{(1)} d\Omega =\ {\alpha N_c\over 2\pi^2}\ {x_{01}^2 d^2 x_2\over x_{12}^2\ x_{02}^2}\ dy_2
\end{equation}

\noindent as the probability of emitting a soft gluon, and where we have defined $dy_2= dz_2/z_2.$  Eq.(47) is illustrated in Fig.5.

\subsubsection{Arbitrary numbers of soft gluons}

Once a single soft gluon emission has been calculated it is straightforward to add more, even softer gluons.  The time sequence for soft gluons
to appear in the wavefunction follows their longitudinal momentum.  The hardest of the soft gluons appears first, then the next hardest of the
soft gluons, etc.  In each case the emission of a soft gluon creates a transition from a color dipole to two color dipoles with the probability
factor given by (47).  In order to express this most succinctly it is convenient to introduce a generating functional, Z, obeying

\begin{displaymath}
Z(x_{01}Y, u) = exp\left[-{2\alpha N_c\over \pi}\ Y\ell n({x_{01}\over \rho})\right] u(x_{01}) + {\alpha N_c\over 2\pi^2}
\end{displaymath}
\begin{equation}
{\rm x} \int_R{x_{01}^2 d^2x_2\over x_{02}^2 x_{12}^2} \int_0^Y dy\  exp[-{2\alpha N_c\over \pi} (Y-y) \ell n ({x_{01}\over \rho})] Z(x_{02}, y, u)
Z(x_{21}, y,u).
\end{equation}

\noindent R  is a cutoff and indicates that one take $x_{02} > \rho, x_{12} > \rho$ as the region of ingtegration.  Eq.(48) is illustrated
in Fig.6.   $Z$ is the generating functional for soft gluons and has the following properties:

\begin{center}
\begin{figure}[htb]
\epsfbox{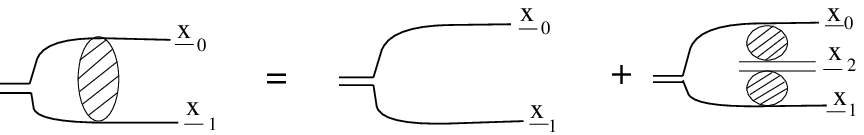}
\end{figure}
\center{FIG.6}
\end{center}

\begin{equation}
{\delta\over \delta u(\Delta x_1)}\ {\delta\over \delta u(\Delta x_2)} \cdot\cdot\cdot {\delta\over \delta u(\Delta x_n)}\ Z\ \Bigg\vert_{u=0}
\end{equation}

\noindent gives, when multiplied by $\Phi^{(0)}(x_{01}, z_1)$ the exclusive probability of having $n - 1$ soft gluons in the onium wavefunction
with the soft gluons and the heavy quark-antiquark pair making up $n$ dipoles of sizes $\Delta x_1, \Delta x_2, \cdot\cdot\cdot .$  If one takes
$u=1,$ rather than $u=0,$ the expression (49) becomes the inclusive probability of having $n-1$ measured soft gluons, along with an arbitrary
number of unmeasured gluons, in the onium wavefunction.  Finally

\begin{equation}
Z(x_{01}, Y, u) \bigg\vert_{u=1}=1
\end{equation}

\noindent is probability conservation and is obtained by including the probability conserving exponential factors in (48) which correspond to
virtual loop corrections in the wavefunction\cite{Z.Ch}.

\begin{center}
\begin{figure}[htb]
\epsfbox{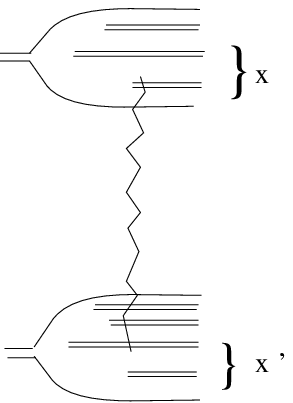}
\end{figure}
\center{FIG.7}
\end{center}

\noindent{\bf Problem 13(H):}  Show that the virtual corrections, the terms in the exponentials in (48) can be obtained using probability
conservation by integrating single gluon emission over the available (cutoff) phase space.  That is
\vskip15pt
\begin{displaymath}
{\alpha N_c\over 2\pi^2}\ \int_R\ {x_{01}^2d^2x_2\over x_{02} x_{12}^2} \int_0^Y dy = {2\alpha N_c\over \pi} Y\ell n\ {x_{01}\over \rho}\ +
O(\rho^2).
\end{displaymath}
\vskip15pt

In summary (48) represents the soft gluon wavefunction of a high energy heavy onium.  The longitudinal momentum integrations have been done in
a leading logarithmic approximation, while the transverse integrations have been handled exactly.  The cutoff $\rho$ will disappear when a
physical problem is considered.  The large $N_c$ limit is esential to obtain (48).

\subsubsection{Scattering in the BFKL approximation}

In terms of the heavy onia light-cone wavefunctions, high energy onium-onium scattering is very simple and is reminiscent of partonic
expressions in hard scattering.  Suppose the scattering takes place in the center of mass.

Then the scattering proceeds by the interaction (scattering) of a single dipole in the left-moving onium with a single dipole in the
right-moving onium as illustrated in Fig.7.  The equation is

\begin{equation}
\sigma(Y) = \int {d^2xd^2x^\prime\over 4\pi^2x^2(x^\prime)^2} N(x, {Y\over 2}) N(x^\prime, {Y\over 2}) \sigma_{dd}(x,x^\prime),
\end{equation}

\noindent where $N(x,Y)$ is the number density of dipoles of size $x$, in a rapidity interval $Y,$ in the onium wavefunction.  More
explicitly

\begin{equation} 
N(x,Y) =  \int d^2 x_{01} dz_1 \Phi^{(0)}(x_{01},z_1) n (x_{01}, x,Y)
\end{equation}

\noindent with

\begin{equation}
n(x_{01}, x,Y) = x^2 \int d\phi (\underline{x})\ {\delta Z(x_{01},Y,u)\over \delta u(x)}\bigg\vert_{u=1},
\end{equation}

\noindent where $d\phi(\underline{x})$ indicates an integration over the orientations of the dipole direction $\underline{x}.$  $n$ obeys the
dipole version of the BFKL equation

\begin{displaymath}
n(x_{01}, x,Y) = x\delta(x - x_{01}) exp\left[ - {2\alpha N_c\over \pi}Y\ell n {x_{01}\over \rho}\right] + {\alpha N_c\over \pi^2} \int_R {x_{01}^2
d^2x_2\over x_{12}^2 x_{02}^2}
\end{displaymath}
\begin{equation}
\times  \int_0^Y dy\  exp\left[-{2\alpha N_c\over \pi}(Y-y)\ell n{x_{01}\over \rho}\right]n(x_{12}, x,y)
\end{equation}

\noindent illustrated in Fig.8
\begin{center}
\begin{figure}[htb]
\epsfbox{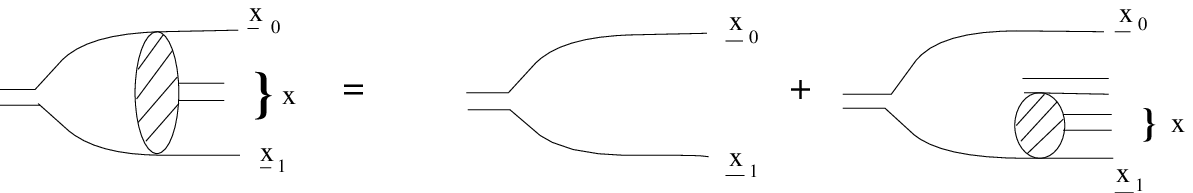}
\end{figure}
\center{FIG.8}
\end{center}

\noindent{\bf Problem 14(M):} Use (48) and (53) to derive (54).

In order to solve (54) it is convenient to write  it as an evolution equation in Y.  Taking a Y-derivative on both sides of (54) gives

\begin{equation}
{d\over dY} n(x_{01},x,Y) = {2\alpha N_c\over \pi} \int_0^\infty dx_{12}^\prime K(x_{01},x_{12}^\prime) n (x_{12}^\prime,x,Y)
\end{equation}

\noindent with

\begin{equation}
K(x_{01},x_{12}^\prime)= {1\over 2\pi} \int_R d^2x_2\delta(x_{12}-x_{12}^\prime)\left\{{x_{01}^2\over x_{12}^2\  x_{02}^2}\ -
2\pi\delta(\underline{x}_2-\underline{x}_0) \ell n({x_{01}\over \rho})\right\}
\end{equation}

\noindent The limit $\rho \to 0$ can be taken in (56) leaving \ K\ a scale covariant kernel.  Thus

\begin{equation}
\int dx_{12}K(x_{01}, x_{12}) x_{12}^{1+2i\nu} = \chi(\nu )x_{01}^{1+2i\nu}
\end{equation}

\noindent from the scale covariance.  Explicit calculation[6] gives

\begin{equation}
\chi(\nu) = \psi(1) - {1\over 2} \psi({1\over 2} + i\nu) - {1\over 2}\psi({1\over 2}-i\nu),
\end{equation}

\noindent thus proving that \ K\ is the usual BFKL kernel in a different guise.  Using (57) in (55) gives

\begin{equation}
n(x_{01},x,Y) = c \int_{-\infty}^\infty {d\nu\over 2\pi}({x_{01}\over x})^{1+2i\nu} e^{{2\alpha\ N_c\over \pi}\chi(\nu)Y}.
\end{equation}

\vskip 15pt
\noindent{\bf Problem 15(E):} Use\ $n(x_{01},x,0) = x\delta(x_{01}-x)$ to how that $c = 2$ in (59).
\vskip 15pt
\noindent{\bf Problem16(M):} Use the fact that $\nu  = 0$ is a saddle point of $\chi(\nu) ,$ along 
with the results $\chi(0) = 2 \ell n$ and
$\chi^{\prime\prime}=\ -\ 14\zeta(3)$ to get

\begin{equation}
n(x_{01},x,Y) = {x_{01}\over 2x} {e^{(\alpha_P-1)Y}\over {\sqrt{{7\over 2}\alpha N_c\zeta(3)Y}}} exp\left[-{\pi \ell n^2({x_{01}\over x})\over
14\alpha N_c\zeta(3)Y}\right].
\end{equation}

\noindent Finally, using (52) and (60) in (51) gives

\begin{equation}
\sigma(Y) = 16\pi\alpha^2R^2 {e^{(\alpha_P-1)Y}\over {\sqrt{{7\over 2}\alpha N_c\zeta(3)Y}}}
\end{equation}

\noindent with

\begin{equation}
\alpha_P - 1 = {4\alpha N_c\over \pi}
\ell n 2
\end{equation}

\noindent and

\begin{equation}
R = {1\over 2} \int d^2x_{01} \int_0^1 dz_1x_{01}\Phi^{(0)}(x_{01},z_1).
\end{equation}

\section{Phenomenology}

In this chapter we are going to try and confront the ideas of quark and gluon saturation with data.  Before doing this we need to further
refine the idea of saturation and relate it more precisely with that of unitarity.  We begin our discussion by turning to soft hadronic
interactions where the concept of parton does not apply but where there is significant evidence that cross sections have reached their unitary
limit for small impact parameter scattering.

\subsection{Soft hadron-hadron scattering[14,15]}

Total and elastic cross sections are well represented in the Regge picture as

\begin{equation}
\sigma_{tot}=c_1(s/s_0)^\epsilon + c_2\bigg/{\sqrt{s/s_0}}
\end{equation}

\noindent with $\epsilon \approx 0.1$ and the constants $c_1$ and $c_2$ depending on the particular scattering being considered.  $\epsilon,$
however, appears to be universal.  There is a major difficulty with a simple \underline{Regge pole} picture, though, in that diffractive cross
sections in proton-antiproton collisions grow much more slowly than expected\cite{K.Go}.  The understanding of why diffractive cross sections
grow more slowly at Tevatron energies than at lower energies is most easily seen in an impact parameter picture of high energy
scattering\cite{U.Am}.   One writes the total, inelastic and elastic cross sections in terms of the S-matrix at an impact parameter\ b\ as

\begin{eqnarray}
 \nonumber\sigma_{tot} &=& 2\int d^2b [1-S(b)]\\
\sigma_{in} &=& \int d^2b [1-S^2(b)]\\
\nonumber\sigma_{e\ell} &=& \int d^2b [1-S(b)]^2.
\end{eqnarray}

\begin{center}
\begin{figure}[htb]
\epsfbox{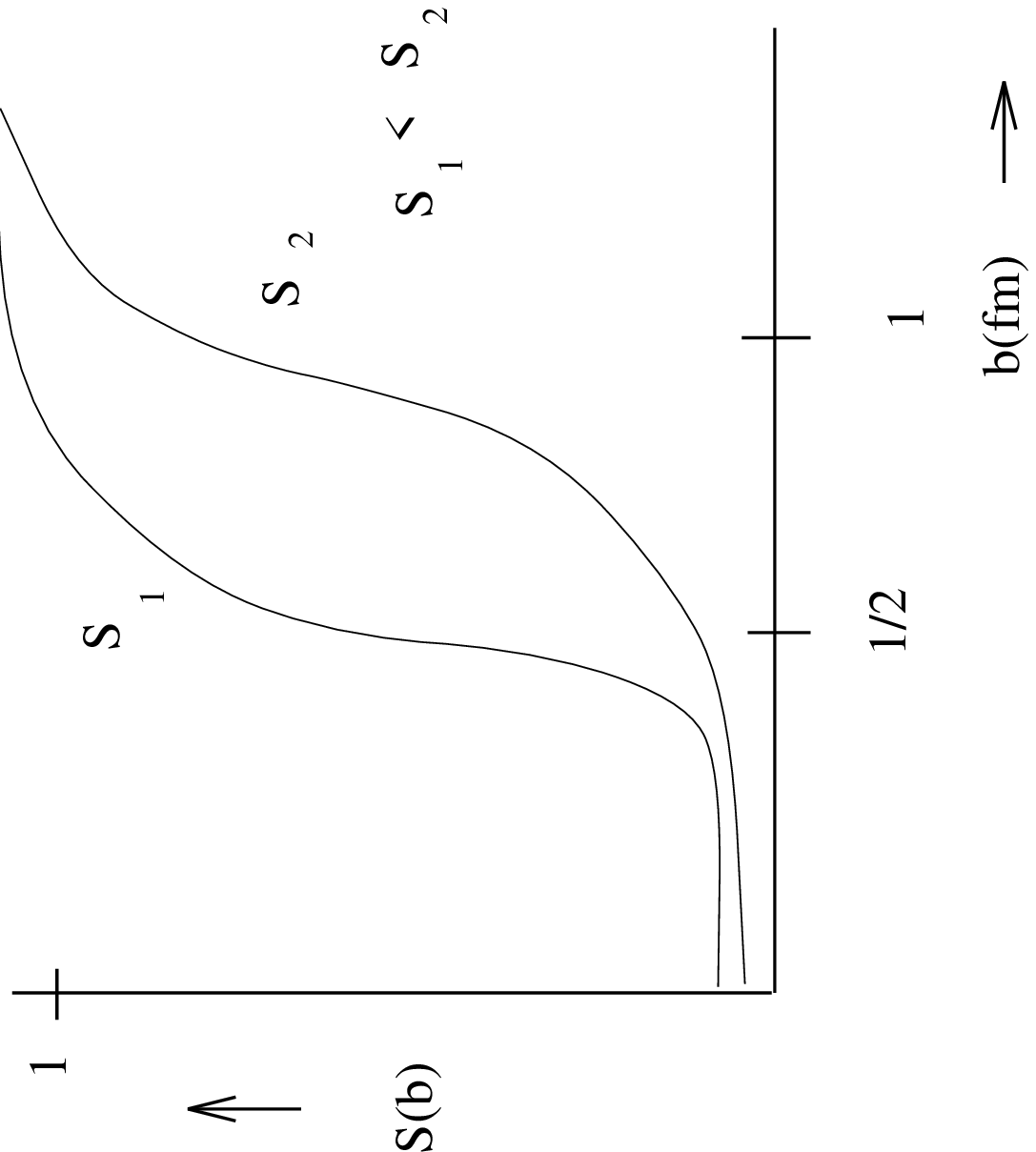}
\end{figure}
\center{FIG.9}
\end{center}

\noindent Equation (65) assumes that $S(b)$ is real and it also neglects internal degrees of freedom in the proton relying on some average
configuration in the proton as dominant in the scatterings described in (65).  Using (65) to fit proton-proton and proton-antiproton
scattering leads to a picture of $S(b)$ as (schematically) illustrated in Fig.9. What is interesting in Fig.9 is that the S-matrix is completely
black, that is unitarity limits have been reached, for significant regions of $b$ in high  energy proton-antiproton collisions.  In the
regions which  are completely black $(S=0)$ elastic and elastic cross sections are equal.  Diffractive production can occur in the region where
$S$ goes from $0$ to $1$ but should not occur either when $S$ is very near zero or when $S$ is close to one\cite{L.L.F}.  The increase in
$\sigma_{tot}$ with energy occurs because some regions in $b$ are changing from grey to black and because  regions of larger  $b$ are going
from white to grey.  The region which contributes to diffraction grows less rapidly than those regions in $b$ giving elastic and highly
inelastic scattering.  The Regge pole picture can be expected to be valid only when
$S$ is not too small.  The regions where  $S$ is small correspond to multiple pomeron exchange in a Regge picture\cite{A.Ca,A.B.K}.

\subsection{Saturation of parton densities and unitarity}
\subsubsection{Two useful reference frames in deep inealstic scattering}

Different frames of reference can be useful in understanding specific properties of deep inelastic lepton-proton and lepton-nucleus scattering.
For example, the Bjorken frame makes the view of the virtual photon as a probe of structure in the proton very natural.  This picture of deep
inelastic scattering is illustrated in Fig.10 where $p$ is very large at small x
\begin{center}
\begin{figure}[htb]
\epsfbox{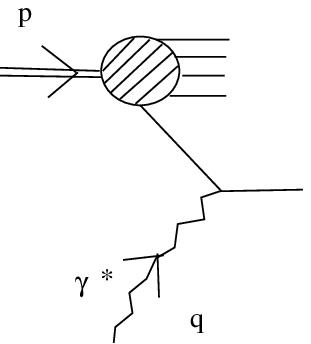}
\end{figure}
\center{FIG.10}
\end{center}

\begin{eqnarray}
\nonumber &p\approx& (p + {m^2\over 2p}, 0,0,p)\\
 &q =& (0,0,0,-Q).
\end{eqnarray}

\noindent The Bjorken variable $x$ is equal to ${Q\over 2p}$ in this frame which is close to the original Bjorken frame except that $q_\perp
=0$ so that the fact that the $\gamma^\ast$ is a local probe in $x_\perp$ at large \ $Q$\  is not manifest.  However, in this frame the virtual
photon can still be viewed as being absorbed by a quark over a short period of time thus making manifest that $F_2$ can be described in terms
of the number of quarks in the proton.  The frame just described, a Bjorken-like or Breit-like frame, is connected by a simple boost to
another frame, which we shall call the dipole frame, where the process appears in time to proceed by the virtual photon breaking up into a
quark-antiquark pair which then scatters on the proton (or nucleus).  In this frame, illustrated in Fig.11, the momenta are

\begin{center}
\begin{figure}[htb]
\epsfbox{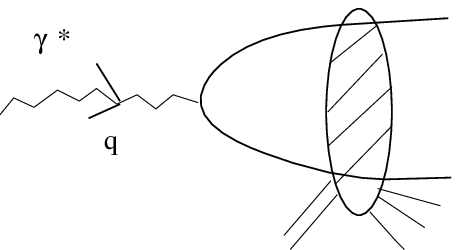}
\center{FIG.11}
\end{figure}
\end{center}

\begin{eqnarray}
\nonumber&p\approx&(p+ {m^2\over 2p}, 0,0,p)\\
&q=& ({\sqrt{q^2-Q^2}}0,0,-q) 
\end{eqnarray}

\noindent with $x = {Q^2\over 2p[{\sqrt{q^2-Q^2}} + q]}$ and where we choose $q/Q$ greater than one but not too large.  The idea is to have the
virtual photon break up into a quark-antiquark pair but to keep all the structure and evolution in the proton rather than in a further
evolution of  the quark-antiquark pair.  This will be the case if $q/Q$ is of moderate size.  In this dipole frame the parton picture of deep
inelastic scattering is not  manifest, however, unitarity constraints are manifest\cite{Gri,M.St,ller}.  In particular when a virtual photon
breaks up into a quark-antiquark pair having a definite transverse coordinate separation, $\underline{x},$ and at a definite impact parameter,
$\underline{b},$ that pair scatters off the proton or nucleus with an S-matrix element $S(\underline{b},\underline{x})$ which obeys $\vert S(
\underline{b}, \underline{x})\vert < 1.$  We shall see that $S(\underline{b}, \underline{x}) = 0$ corresponds to saturation.  Thus
saturation (Bjorken frame) corresponds to blackness of the cross section (dipole frame).

\subsubsection{Scattering off a nucleus (simple picture)}

To illustrate how quark saturation comes about consider deep inelastic scattering off a large nucleus in the dipole frame.  We simplify the
problem by only allowing the individual nucleons of the nucleus to interact with the quark-antiquark pair either through one gluon exchange, an
inelastic reaction, or through two gluon exchange.  We do not allow evolution in the individual nucleons so one may view this model as an
extension of the McLerran-Venugopalan model where now we have a (quantum) quark-antiquark pair interacting with the quasiclassical gluon
field\cite{ller,W.Bu,Geh,Dos,Zak,For} of the nucleus which we calculated in Chapter 1.  However, since the nucleons in the nucleus do not have
any evolution we can also simply take the nucleus at rest in which case the cross section takes the form of a virtual photon breaking into a
quark-antiquark pair which then multiply scatters, via one and two gluon exchange, with the nucleons in the nucleus.  This view is illustrated
in Fig.12.

\begin{center}
\begin{figure}[htb]
\epsfbox{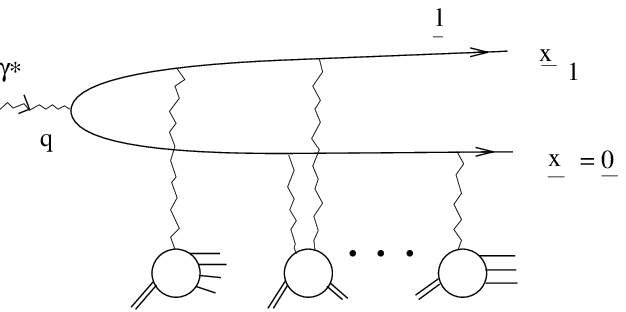}
\center{FIG.12}
\end{figure}

\end{center}

Using

\begin{equation}
F_2(x, Q^2) = {1\over 4\pi^2\alpha_{em}} Q^2(\sigma_T+\sigma_L)
\end{equation}

\noindent and

\begin{equation}
F_2(x, Q^2) = \sum_f e_f^2[x q_f(x,Q^2) +  x \bar{q}_f(x, Q^2)]
\end{equation}

\noindent and the fact that $\sigma_L/\sigma_T << 1 \ (\sigma_T\ \  {\rm and}\ \ \sigma_L$ are the cross sections for transverse and longitudinal
photons, respectively) one can directly write a a formula for the quark and antiquark distributions in terms of the cross section for
virtual photon scattering on the nucleus.  In fact, we wish to go somewhat beyond this and write a formula for the measurement of a leading
(current) jet in the deep inelastic scattering The equation is\cite{ller}

\begin{displaymath}
e_f^2{dxq_f\over d^2\ell d^2b}\ = {Q^2\over 4\pi^2\alpha_{em}} \int {d^2x_1d^2x_2\over 4\pi^2} \int_0^1 dz{1\over2}
\sum_\lambda \psi_{T\lambda}^{^\ast f}(\underline{x}_2,z)\psi_{T\lambda}^f(\underline{x}_1,z)
e^{-i\underline{\ell}\cdot(\underline{x}_1-\underline{x}_2)}
\end{displaymath}
\begin{equation}
\left[1+e^{-(\underline{x}_1-\underline{x}_2)^2\bar{Q}_s^2/4} - e^{\underline{x}_1^2\bar{Q}_s^2/4}-e^{-\underline{x}_2^2\bar{Q}_s^2/4}\right]
\end{equation}

\noindent with the photon wavefunction into a quark-antiquark pair of flavor   $f$

\begin{equation}
\psi_{T\lambda}^f(\underline{x},z) =      [{\alpha_{em}N_c\over 2\pi^2}\ z(1-z)[z^2+(1-z)^2]Q^2      ]^{1/2} e_f
K_1 ({sqrt{Q^2\underline{x}^2z(1-z)}}) {\underline{\epsilon}^\lambda\cdot\underline{x}\over \vert\underline{x}\vert},
\end{equation}

\noindent and where the quark saturation momentum, $\bar{Q}_s,$ obeys

\begin{equation}
\bar{Q}_s^2 = {C_F\over N_c} Q_s^2.
\end{equation}

\noindent the various factors in (70) have a ready explanation.  The factor of ${Q^2\over 4\pi^2\alpha_{em}}$ comes from (68) relating a
structure function (or parton distribution) to a virtual photon cross section.  The wavefunctions give the amplitude, and complex conjugate
amplitude, for a virtual photon to go into a quark-antiquark pair with momentum fractions $z$ and $1-z$ and at transverse coordinate
separation $\underline{x}_1$ in the amplitude and $\underline{x}_2$ in the complex conjugate amplitude.  The integration over $d^2x_1$ and
$d^2x_2$ along with the 
$e^{-i\underline{\ell}\cdot(\underline{x}_1-\underline{x}_2)}$ factor fix the transverse momentum of the quark (or
antiquark) to be $\underline{\ell}.$  The factors in the square bracket represent (respectively) no scattering of the pair, scattering in the
amplitude and complex conjugate amplitude, scattering in the amplitude, and scattering in the complex conjugate amplitude.  The exponential
factors in the  [ ] are recognizable as multiple scatterings if one identifies

\begin{equation}
{\bar{Q}_s^2\underline{x}^2\over 4} = {2\sqrt{R^2-b^2}} \rho {\sigma(\underline{x})\over 2}
\end{equation}

\noindent with $\rho$ the usual nucleon density and $\sigma$ the  cross section of a quark-antiquark
dipole of separation $\underline{x}$ on a nucleon.  The ${1\over 2}$ in front of the $\Sigma$ in (70) is for normalization and means we are
describing the jet rate for a single transversely polarized virtual photon.

If $\underline{\ell}^2$  is much less than $\bar{Q}_s^2$ the integrals in (70) can be done giving

\begin{equation}
{dxq_f\over d^2\ell d^2b} = {N_c\over 2\pi^4}
\end{equation}

\noindent which says there is ${1\over 2\pi^4}$ quarks, or antiquarks, per unit of transverse phase space in the wavefunction of the nucleus. 
This is quark saturation.  This result follows from the unitarity of the dipole-nucleus scattering.

If one integrates (70) over $\underline{\ell}$\  and\  $\underline{b}$, the result

\begin{displaymath}
 x q_f(x,Q^2) + x \bar{q}_f(x,Q^2) = {Q^2N_c\over 8\pi^4}\int d^2b\int d^2x \int_0^1 dz\  z(1-z)[z^2 + (1-z)^2]
\end{displaymath}
\begin{equation}
K_1^2[{\sqrt{Q^2\underline{x}^2z(1-z)}} ](1-e^{-\underline{x}^2\bar{Q}_s^2/4})
\end{equation}

\noindent emerges.  We have in fact reached what is essentially the Golec-Biernat and W\"usthoff model\cite{Gol} a model of saturation
successfully applied to both structure functions and diffractive producton.  The identification is complete if we identify $\bar{Q}_s =
1/R_0(x)$\ and\ $\int d^2b = \sigma_0.$  Golec-Biernat and W\"usthoff do not do their analysis in impact parameter space but introduce a
parameter setting the scale for cross  sections, $\sigma_0,$ which is about 25 mb in their fits.  They also parameterize $R_0^2(x) = {1\over
GeV^2}({x\over x_0})^\lambda$ with $\lambda \approx 0.3$\ and $x_0 \approx 3x10^{-4}.$  The fact that they are able to get such good fits of
the data at low and  moderate $Q^2$ for $F_2$ and for diffractive structure functions is, perhaps, the strongest evidence that saturation
has been reached in the HERA energy range.

Although (70) and (75) have been derived for scattering on a large nucleus the picture would seem to be pretty general.  All the QCD dynamics
is in the four factors in the [ ] in (70).  The identification of the exponential with the S-matrix according to

\begin{equation}
S(\underline{b}, \underline{x}) = e^{-\underline{x}^2Q_s^2/4}
\end{equation}

\noindent means we may everywhere replace the gaussian exponential factors by the S-matrix and have a very general description of deep
inelastic scattering, on either a proton or a nucleus, where unitarity is controlled.  The essential physics which the Golec-Biernat and
W\"usthoff model incorporates is

\begin{equation}
S(\b{b}, \underline{x}) = 1\  {\rm when}\ \underline{x}^2\bar{Q}_s^2/4 <<1
\end{equation}

\begin{equation}
S(\underline{b}, \underline{x}) = 0 \ {\rm when}\ \underline{x}^2\bar{Q}_s^2/4 >>1
\end{equation}

\noindent for some scale $\bar{Q}_s$ along with the statement that the $b$ dependence is on the scale of a fermi for protons and several fermis
for nuclei.

\subsubsection{Is there evidence for saturation?}

Although the Golec-Biernat and W\"usthoff model gives the strongest evidence for saturation I would like to go over the argument in a very
qualitative way because the issue is so important and because the physics argument, though not technically difficult, is subtle.  In general
fits to $F_2(x, Q^2)$ using a second order DGLAP formalism work very well in the HERA regime.  However, there are signs that all is not well
with such fits in the region of very small x and moderate values of $Q^2.$  In addition there is evidence that something like saturation is
needed in order to be consistent with the energy dependence of diffractive structure functions measured at HERA.

We begin with $F_2(x, Q^2).$  Sometime ago A. Caldwell\cite{Cal,Abr} presented evidence for a change in the $Q^2-$ dependence of $F_2$ at very
small \ $x$\ and moderate $Q^2.$  The HERA data is shown in Fig.13 where an unexpected turnover in ${\partial F_2\over \partial \ell n Q^2}$
occurs at moderate $Q^2$ and small $x.$  Data through a similar region of $Q^2,$ but at much larger values of $x$ from fixed target experiments
are shown in Fig.14 where there is no turnover visible at moderate $Q^2.$  This turnover is well described by a formula like that given in
(75), that is by the model of Golec-Biernat and W\"usthoff.  Let us here try to describe what may be happening in very simple
terms\cite{Core,Mao}.

\vspace{.4cm}
\begin{center}
\epsfxsize=10cm
\epsfysize=10cm
\epsfbox{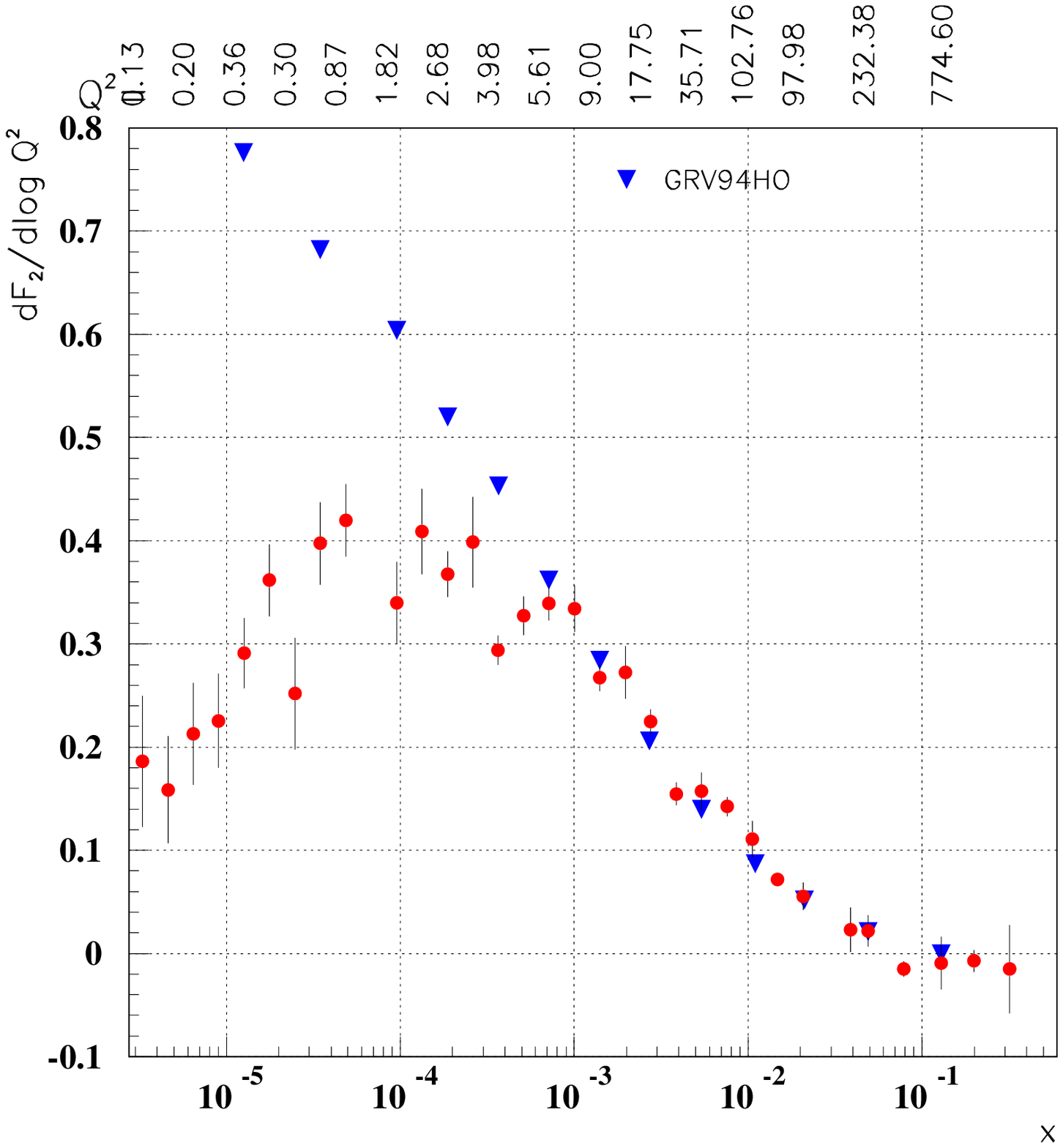}
\center{FIG.13}

\end{center}

    In the dipole frame the virtual photon breaks up into a quark-antiquark pair of relative momentum $2\underline{k}$ which then scatters,
inelastically, on the proton as illustrated in Fig.15.  We may write

\begin{figure}[htb]
\epsfxsize=10cm
\epsfysize=10cm
\epsfbox{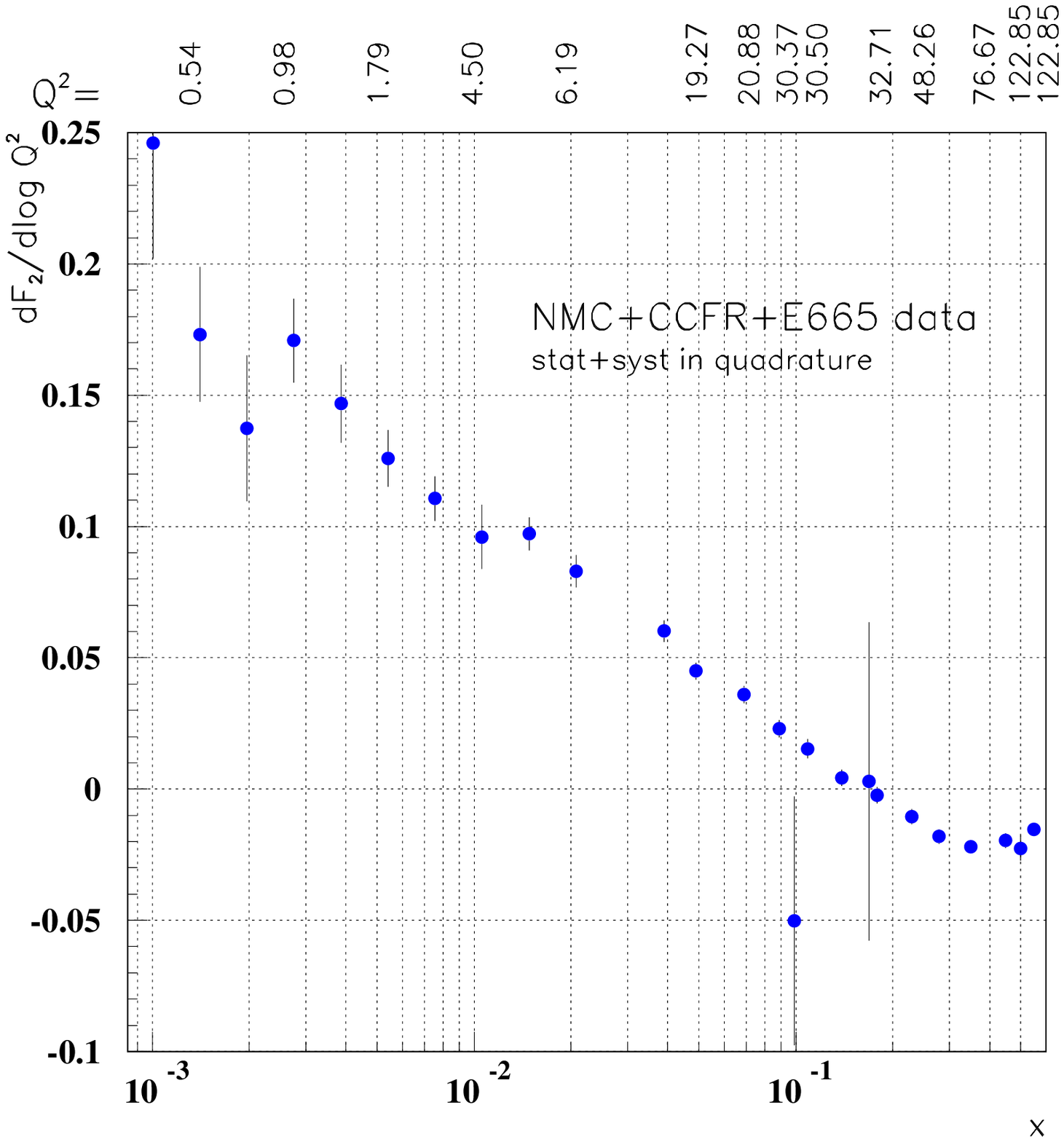}
\center{FIG.14}
\end{figure}

\begin{figure}[htb]
\epsfbox{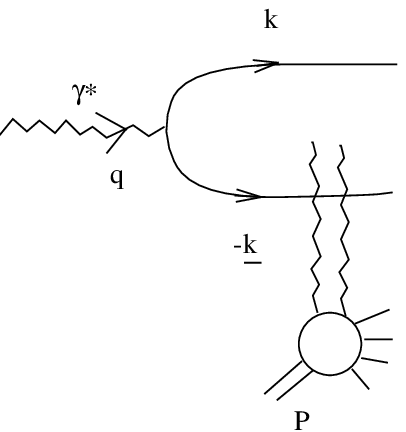}
\center{FIG.15}
\end{figure}

\begin{equation}
dF_2 \propto Q^2 dP_r(\underline{k}^2) \sigma_{in}(\underline{k})
\end{equation}

\noindent where

\begin{equation}
dP_r(\underline{k}^2) \propto {d\underline{k}^2\over Q^2}
\end{equation}

\noindent is the probability that the virtual photon break up into the quark-antiquark pair in the momentum range $d\underline{k}^2.\  
\sigma_{in}$ is the cross section for the pair to interact with the photon.  In perturbation theory

\begin{equation}
\sigma_{in}^{Pert}(\underline{k}) \propto {\alpha\over \underline{k}^2} x G(x, \underline{k}^2)
\end{equation}

\noindent so that $Q^2$ cancels out in (79) and one is left with the usual $d\underline{k}^2 \underline{k}^2$ integral giving scaling
violations.  In this case $Q^2{\partial F_2\over \partial Q^2}$\   $\alpha x G(x, Q^2)$ and one would interpret the turnover in the
Caldwell plot as a (surprising) decrease in the gluon distribution at very low values of $x.$  However, if $\underline{k}^2$ is in the
saturation regime

\begin{equation}
\sigma_{in}^{sat}(\underline{k}) = \pi R_0^2,
\end{equation}

\noindent with $R_0$ about the proton's radius, since the unitarity limit is being reached.  This means that $Q^2 {\partial F_2\over \partial
Q^2}\ \sim Q^2R_0^2$ and the turnover at small $Q^2$ is seen as a natural result of unitarity, or saturation.  It would be expected that
unitarity, and saturation, should occur earlier for small impact parameter interactions than for large impact parameter interactions.  It
would be very interesting to be able to do analysis of the $Q^2-$ dependence of structure functions at very small $x$ as a function of the
impact parameter of the collision.

The other process for which there is some evidence for saturation is in the diffractive cross section at HERA\cite{Gol,vin,Bar,Kow}.  The
picture again is as in Fig.15 but where now the proton, p, must exchange at least two gluons, and remain in a color singlet, in order that a large
rapidity gap emerge with the proton typically appearing in the final state at a small momentum transfer.  One can write

\begin{equation}
d\sigma_{Dif} \propto dP_r(\underline{k}) \sigma_{e\ell}(\underline{k})
\end{equation}

\noindent with $dP_r$ as in (80).  In perturbation theory

\begin{equation}
\sigma_{el}^{pert} \propto[{\alpha\over \underline{k}^2} x G(x, \underline{k}^2)]^2
\end{equation}

\noindent with $x \approx M^2\big/ s$ and with $M$ the mass of the diffractive state coming from the quark-antiquark pair.  Using (84) in
(83) leads to a quadratically diverging infrared integration $d\underline{k}^2\big/(\underline{k}^2)^2,$ and this suggests that the process of
diffraction at HERA is a soft process.  This was the common wisdom among theorists until a few years ago.  What could happen to make the
process semihard?  If the saturation momentum is as large as a GeV or a few GeV then (84) applies for transverse momenta above the saturation
regime.  If $\vert{\underline{k}\vert}$ is below the saturation regime one should use

\begin{equation}
\sigma_{e\ell}^{sat} \propto \pi R_0^2
\end{equation}

\noindent in (83).  This means that the saturation momentum serves as the infrared cutoff for the ``divergent'' integral when (82) is used in
(81).  If the saturation momentum is in the semihard regime (1-2) GeV then the resulting process is not a soft process and one might expect an
energy dependence which is stronger than that given by the soft pomeron.  In fact the data support a fairly strong energy growth.  If one writes

\begin{equation}
x {d\sigma_{Dif}\over dx} \propto x^{-\gamma}
\end{equation}

\noindent then both ZEUS and H1 find $\gamma$ to be about a factor of 2 larger than that expected from soft pomeron exchange.  What is unique
about the scattering of a virtual photon on a hadron is that the virtual photon wavefunction is dominated by high transverse momentum states,
in contrast to hadrons.  If the unitarity limit, saturation, is reached in some range of transverse momentum the process simply proceeds to a
higher region of transverse momentum.  In diffraction of virtual photons the most important region of transverse momenta in the virtual
photon's wavefunction should be just above the saturation momentum.

\end{document}